\documentclass[a4paper,12pt]{article}
\usepackage{graphicx}
\begin{document}

\title{ On a $Z^{\prime}$ signature at next high energy electron-positron colliders.}
\author{ F. M. L. Almeida Jr.\thanks{E-mail: marroqui@.if.ufrj.br}, Y. A.
Coutinho, J. A. Martins Sim\~oes,\\ A. J. Ramalho and S. Wulck. \\ Instituto
de F\'{\i}sica,\\
Universidade Federal do Rio de Janeiro, RJ, Brazil\\
and\\
M. A. B. do Vale\\
Universidade Federal  de S\~ao Jo\~ao del Rei, MG, Brazil}
\maketitle
\begin{abstract}
\par
The  associated production of a $Z^{\prime}$ and a final hard photon in high energy electron-positron colliders is studied. It is shown that the hard photon spectrum contains  useful information on the $Z^{\prime}$ properties. This remark suggests that, if a new neutral gauge boson exists for $M_{Z^{\prime}} < \sqrt{s}$, it will not be necessary to make a new energy run at the $Z^{\prime}$ mass in order to get most of its properties.
 
\end{abstract}
\eject

\section{Introduction}\setcounter{equation}{0}
\par
A general consequence of extended $SU(3)\otimes SU(2)\otimes U(1)$ gauge symmetry is the existence of additional gauge bosons. In the simplest extension one has only a new $U(1)$ neutral gauge boson $Z^{\prime}$. This scheme has a natural place in grand unified theories and in some superstring models \cite{ASS}. In other extended models, like left-right $SU_{L}(2) \otimes SU_{R} (2) \otimes U(1)$, there are new charged and neutral gauge bosons \cite{JCP}. No experimental confirmation of these hypotheses has been found yet, but it is expected that the next generation of colliders will confirm, or rule out, these models. Following the success of the standard  neutral gauge boson $Z$ discovery, it is also expected that a clear sign of the existence of a  
$Z^{\prime}$ boson could be found in the resonant production of lepton pairs via $ Z^{\prime} \longrightarrow \ell^+ \ell^- $ ( with $ \ell = e,\mu ,\tau$ ), in next high-energy colliders \cite{CVE, GCV, DJO}. These topics have been studied by many authors over the last years \cite{ DJO,NLC,PDG}.
\par
However, in the search for this new neutral gauge boson, there is a major difference with the standard model search for the usual $Z$.  As the $Z$ mass was theoretically predicted, colliders were built at the $Z$ mass energies necessary to study its properties. Since the $Z^{\prime}$ mass is not known, colliders are considered at the highest feasible energy $ \sqrt s$. If indications of a $Z^{\prime}$ signal are found with a mass lower than $ \sqrt s $, a new run near the mass value would be necessary in order to study the $Z^{\prime}$ properties in detail. This could be experimentally a very complex, expensive or even an impossible operation. It is important to study alternative methods that could equally well disentangle the $Z^{\prime}$ properties without changing the collider energy. The soft-photon emission accompanying a new $Z^{\prime}$ is known \cite{APE} to imply logarithm corrections to the cross section. For the standard neutral gauge boson production associated with a neutrino pair it was noted \cite{CAR} that the $Z$ pole change its position. Recently \cite{FRE} a study was presented of the process $ e^+ e^-   \longrightarrow  Z^{\prime}
 n \gamma$ . The effect of the multi $\gamma$ emission  is to reduce the effective available energy and the consequent production of a real $Z'$ with a mass below $\sqrt s $.
\par
The purpose of this letter is to present an alternative signature for $Z^{\prime}$ production at the new electron-positron colliders, that could allow us to study its width, decay channels, couplings and so on, at a fixed collider energy $\sqrt s > M_{Z^{\prime}}$.

\section{The Model}\setcounter{equation}{0}
\par
Our main point is the associated production of a $Z^{\prime}$ and a hard photon in the process

\begin{equation}
  e^+ e^- \longrightarrow Z^{\prime} \gamma.
\end{equation}
\par

 This process is similar to the proposal of ref. \cite{APE,CAR,FRE} but with an 
important difference. Whereas in references \cite {APE,CAR,FRE} it is studied multi-soft photon 
emission and the $Z^{\prime}$ properties are obtained from its direct decay products, in our proposal one has a single hard photon emission and the $Z^{\prime}$ properties are studied from this hard photon. A similar proposal for  two body process, with one light particle and another very heavy in the final state , was already studied \cite{NOS} for the case of two fermion production. In the present paper we discuss several advantages of the direct study of the hard photon emission. 
\par
A very simple consequence of  four-momentum conservation of process (1) is that the final high-energy hard photon has an energy given by

\begin{equation}
  E_{\gamma}\mp \Delta_{\gamma}=\frac{s -({M_{Z^{\prime}}\pm \Delta_{Z^{\prime}} })^2}{2 \sqrt s},
\end{equation}
where $\Delta_{\gamma}$ and $\Delta_{Z^{\prime}}$ are the fluctuations in the photon energy $E_{\gamma}$ and $Z^{\prime}$ mass  distributions respectively.
\par

\begin{figure}[t]
       \begin{center}
         \includegraphics[width=6cm,height=6cm]{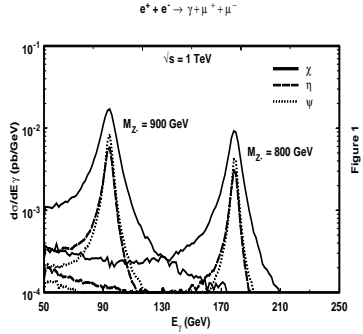}
       \end{center}
\caption{The photon energy distribution in  $ e^+ e^- \longrightarrow  \gamma ~ \mu^- ~ \mu^+ $ for
 $M_{Z^{\prime}}= 800$ GeV and $M_{Z^{\prime}}= 900$ GeV at  $ {\sqrt s}= 1$ TeV for the $ \chi , \eta$ and $\psi$ models. The SM curve is below $10^{-4}$ pb/GeV.
}
\end{figure} 

The study of the hard photon energy distribution  gives the same information as the direct $Z^{ \prime}$ decays, but in a  simple and direct way, without the need to obtain the $Z^{\prime}$ mass from its decay products. In order to obtain numerical estimates, detector and hadronization effects, we will employ the canonical $\eta ,\chi ,\psi $  superstring-inspired $E_6$ models \cite{PDG}, but our arguments apply to any model with extra neutral gauge bosons as well, since it is based on kinematical properties. We neglect  $Z-Z^{\prime}$ mixing and consider that the  
$Z^{\prime}$ couples only to usual fermions. Then there is only one unknown parameter in the above models - the  extra $Z^{\prime}$ mass.
\par

\section{Results}\setcounter{equation}{0}
\par
In order to obtain the hard-photon energy distribution, which is particularly relevant to our analysis, Monte Carlo events were generated and selected by a set of realistic cuts. All final-state particles were required to emerge with a polar angle $\theta$, measured with respect to the direction of the electron beam, in the range $ \mid cos{\,\theta}\mid \le 0.995$. Events in which the hard-photon energy was less than $50$ GeV were ignored. Since we are interested in hard photon emission, this cut eliminates also most of the initial-state radiation. We also imposed a cut $ m_{ij}> 5$ GeV ($i,j=\gamma,\mu^+,\mu^-$) on the invariant masses of the final particles. The $cos{\,\theta}$ and $ m_{ij}$ cuts reflect roughly the detector limitations.
We are assuming that the detector is "blind" for $ \mid cos{\,\theta}\mid \ge 0.995$ and
for cluster with  $ m_{ij}< 5$ GeV.

As an example of future high energy electron-positron colliders we have chosen a new collider  \cite{NLC, NCL} project at an energy  $ {\sqrt s}= 1$ TeV and a typical yearly integrated luminosity of $100$ pb$^{-1}$. 
\par

A first example is given in Fig. 1, which shows the photon energy distribution in the channel $ \gamma \mu^+ \mu^- $ for two $Z^{\prime}$ mass values. In order to account for real and virtual contributions for this process we have included all the twelve Feynman diagrams - eight from the standard model (SM) and four from the extra $Z^{\prime}$ contribution.  We have performed the calculation with the CompHEP package \cite{HEP}. The corresponding distribution for the SM is below 
$10^{-4}$ pb/GeV. A similar distribution, peaked at the photon energy, follows for any other decay channel $ Z^{\prime} \longrightarrow \bar{f} f $. The $ \gamma Z^{\prime}$ total cross section is shown in Fig. 2. The cross section for  the more usual channel $ e^+ e^- \longrightarrow Z^{\prime} \longrightarrow \mu^+ \mu^- $ is also shown for comparison. It is important to observe that the $\gamma Z^{\prime}$ cross section is greater than $\mu^+ \mu^-$ production for all models, including the SM.
\par
In Table 1 we give the $Z^{\prime}$ branching ratios for the fermionic decay channels. For the invisible channel we have summed all neutrinos whereas for the leptonic charged channel individual branching fractions are given.  From this result we can estimate the total number of signal events. 
\par

\begin{figure}[t]
        \begin{center}
         \includegraphics[width=6cm,height=6cm]{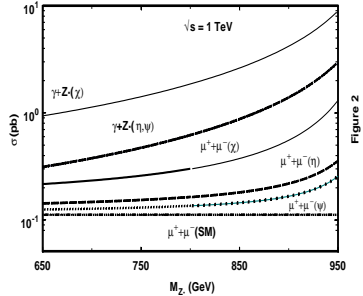}
       \end{center}
\caption{The total cross section for $ e^+ e^- \longrightarrow \gamma ~Z^{\prime}$ and $ e^+ e^- \longrightarrow \mu^+ \mu^-$ for the $ \chi$ , $\eta$, $\psi$ and standard models.
}
\end{figure}

\begin{table}[!t]\label{psiuetachimodels}
\begin{center}
\begin{tabular}{|c|c|c|c|}
\hline \hline
\multicolumn{4}{|c|} {Models} \\ \hline \hline
 Channels &  $\chi$ & $\psi$ & $\eta$  \\ 
\hline Hadrons &  $65,1$ & $79$ & $86$  \\ 
\hline Invisible  & $16,5$ & $6,9$ & $2,4$ \\ 
\hline $l^+ l^-$ & $6,1$  & $4,7$ & $3,9$  \\ 
\hline  $\Gamma_{total}/M_{Z^{\prime}}$  & $0,012$  & $0,005$ & $0,006$  \\ 
\hline 
\end{tabular} 
\end{center}
\caption {$Z^{\prime}$ branching ratios ($\%$) and total width for standard fermions channels in  $\chi$, $\eta$ and $\psi$ models. }
\end{table}

\par
With the purpose of accounting for the finite resolution of the  detectors, we smeared the four-momenta of the final-state photons and leptons by means of the SMEAR routines \cite{SME}. The uncertainties in the energies of the final-state photons were simulated by Gaussian smearing the energies of these particles with a half-width $\Delta E$ of the form $ {\Delta E}/E = a + b/{\sqrt E}$. For the electromagnetic calorimeters proposed for the new linear colliders, $a=1~\%$ and b ranges from $10~\%$ to $15~\%$. We used the value $b=12\%$, which is representative of a  electromagnetic calorimeter \cite{NCL}. The directions of the final-state photons were smeared in a cone around the directions of their original three-momenta, according to a Gaussian distribution with half-width equal to $10$ mrad. As far as the detection of muons is concerned, one has to consider the momentum resolution of the muon tracker, and the multiple scattering effects on the transverse momentum $p_T$ and azimuth $\phi$ of the muons. The details of the procedure to incorporate these effects by Gaussian smearing $1/{p_T}$ and $\phi$ can be found in Settles et al. \cite{SME} and references therein.

\par
\begin{figure}[t]
        \begin{center}
         \includegraphics[width=6cm,height=6cm]{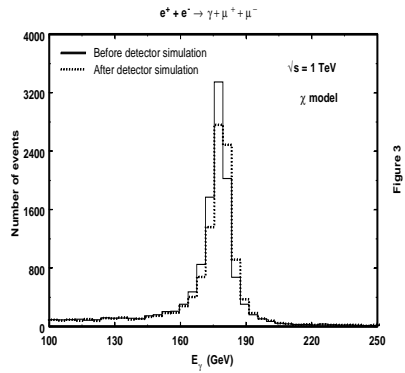}
       \end{center}
\caption{Histogram for the photon energy  before and after detector simulation in  
$ \gamma ~\mu^+ ~\mu^- $ when $M_{Z^{\prime}}= 800$ GeV for $\chi$ model.
}

\end{figure}

The hard photon energy distribution gives a very clear indication of the  $Z^{\prime}$ parameters.  This is shown in Fig. 3 for the channel $ e^+ e^- \longrightarrow Z^{\prime} \longrightarrow  \gamma ~ \mu^+ \mu^- $. The hard photon energy distribution shows practically no difference between the exact theoretical curve and the estimate for the possible data whereas in the reconstructed $ \mu^+ \mu^- $ invariant mass distribution there is a much larger distortion and the peak shifted to left as shown in Fig. 4. This distortion can leads to experimentally sophisticated invariant mass 
correction methods increasing the uncertainties for the $M_{Z^{\prime}}$ and its width.
\par
\begin{figure}[t]
        \begin{center}
         \includegraphics[width=6cm,height=6cm]{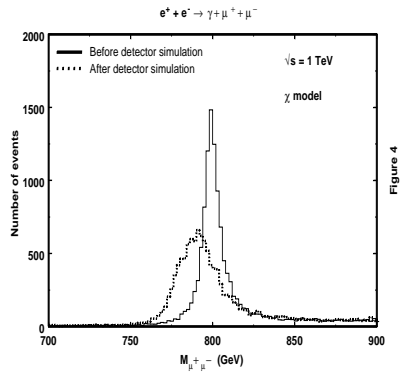}
       \end{center}
\caption{Histogram for the invariant $\mu^- ~\mu^+$ mass before and after detector simulation
when $M_{Z^{\prime}}= 800$ GeV for $\chi$ model.
}
\end{figure}

\begin{figure}[t]
        \begin{center}
         \includegraphics[width=6cm,height=6cm]{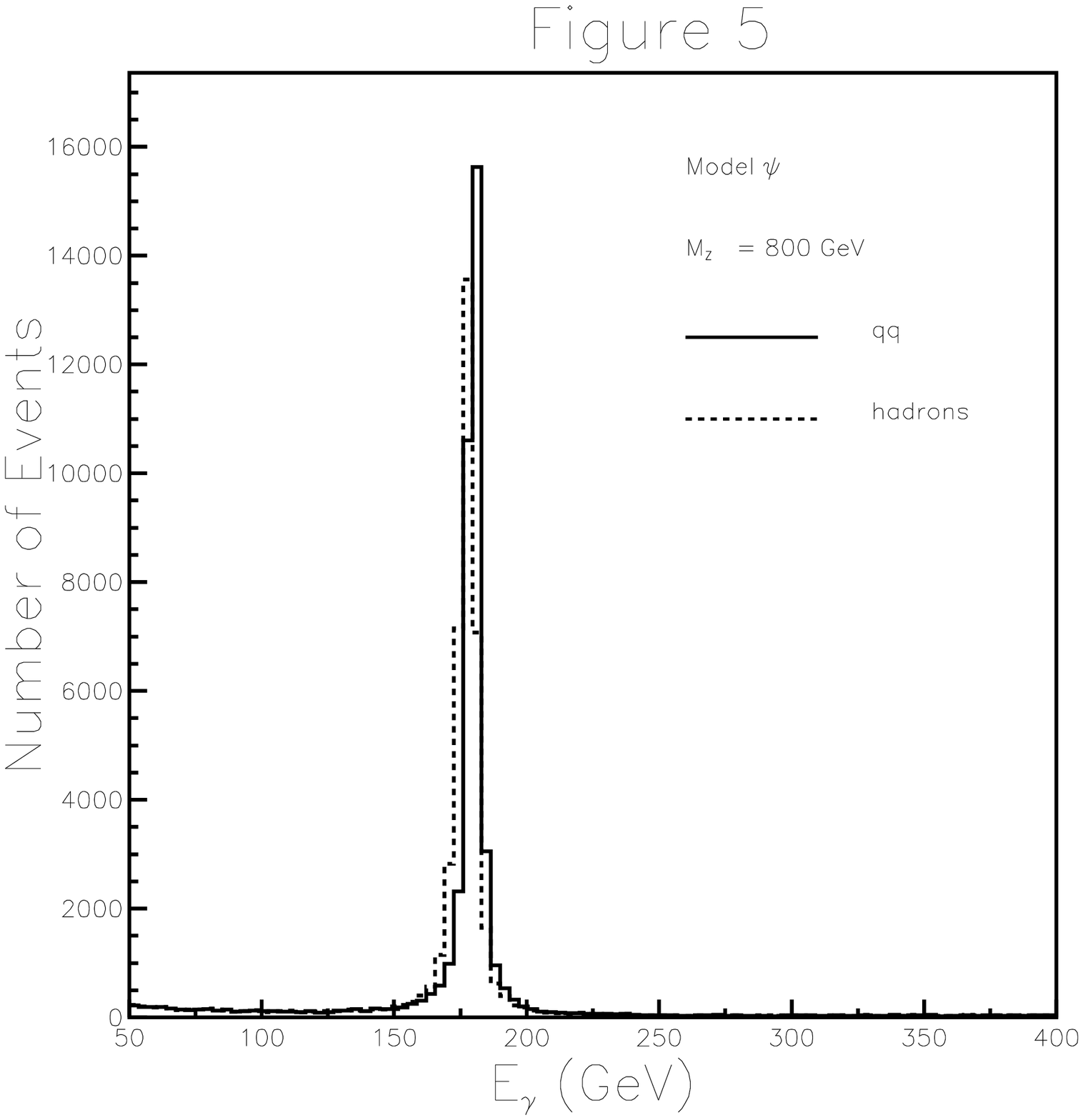}
       \end{center}
\caption{Histogram of the photon energy in $ \gamma ~ jet-jet $ channel before and after hadronization when $M_{Z^{\prime}}= 800$ GeV for $\psi$ model.}
\end{figure}
A similar effect can be seen in the hadronic channels. We have performed the full hadronization procedure for the $Z^{\prime} \longrightarrow  \gamma  ~q  \bar q $ by using the Pythia program. The results are shown in Figs. 5 and 6. Here again the photon peak is practically unchanged, whereas the $jet-jet$ peak presents a much larger distortion. One can also perform the smearing process in the hadronic channels. This will increase even more the distortion in the $Z^{\prime}$ invariant mass reconstruction.

\par
\begin{figure}[t]
        \begin{center}
         \includegraphics[width=6cm,height=6cm]{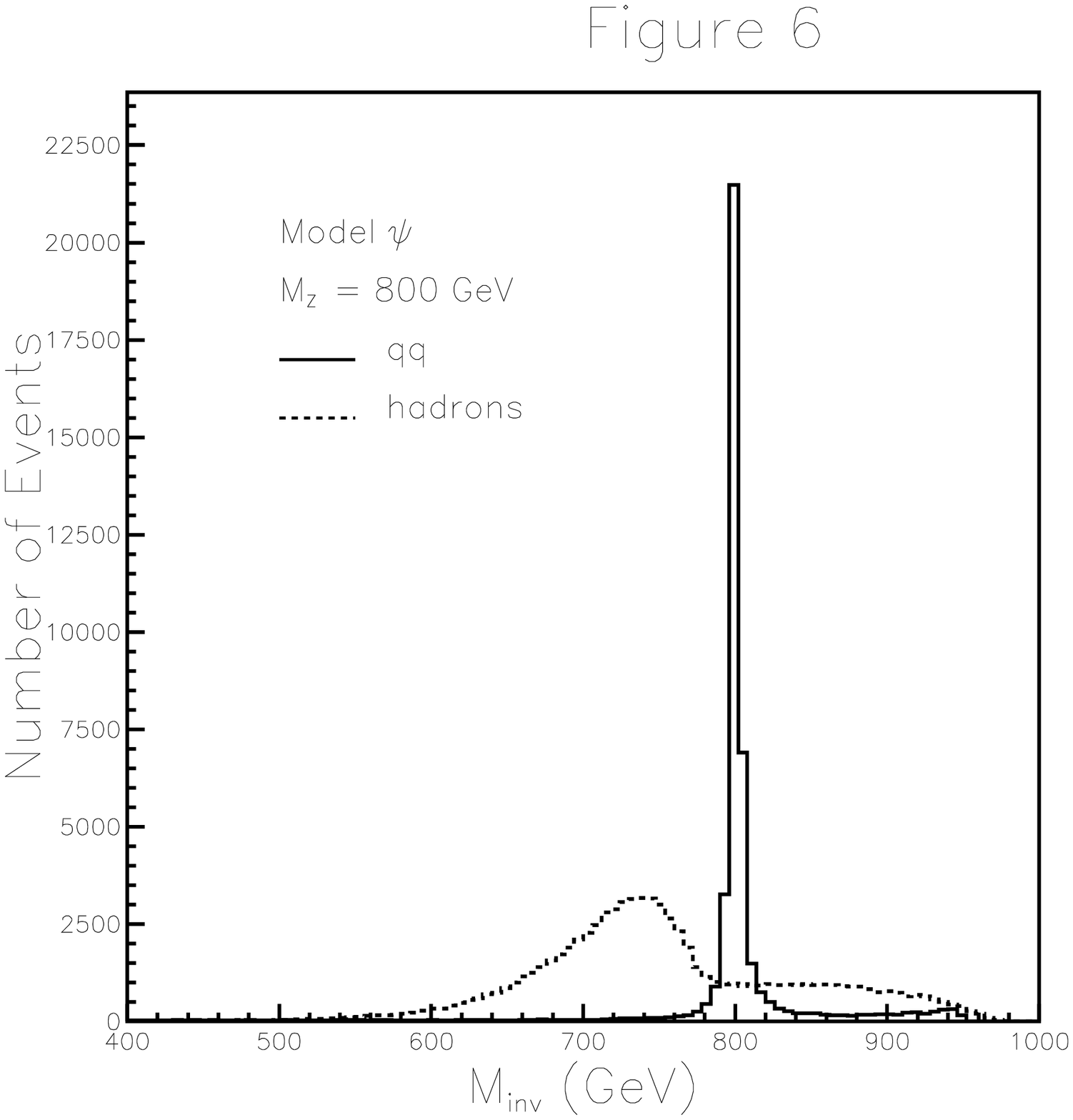}
       \end{center}
\caption{Histogram of the $ jet-jet $ invariant mass before and after hadronization when $M_{Z^{\prime}}= 800$ GeV for $\psi$ model.}
\end{figure}
 In the hard-photon channel one can also study model differences in the charge forward-backward asymmetry, defined relative to the final $ \mu^-$ angular distribution relative to the incoming electron. The result is shown in Fig. 7.
\par\begin{figure}[t]
        \begin{center}
         \includegraphics[width=6cm,height=6cm]{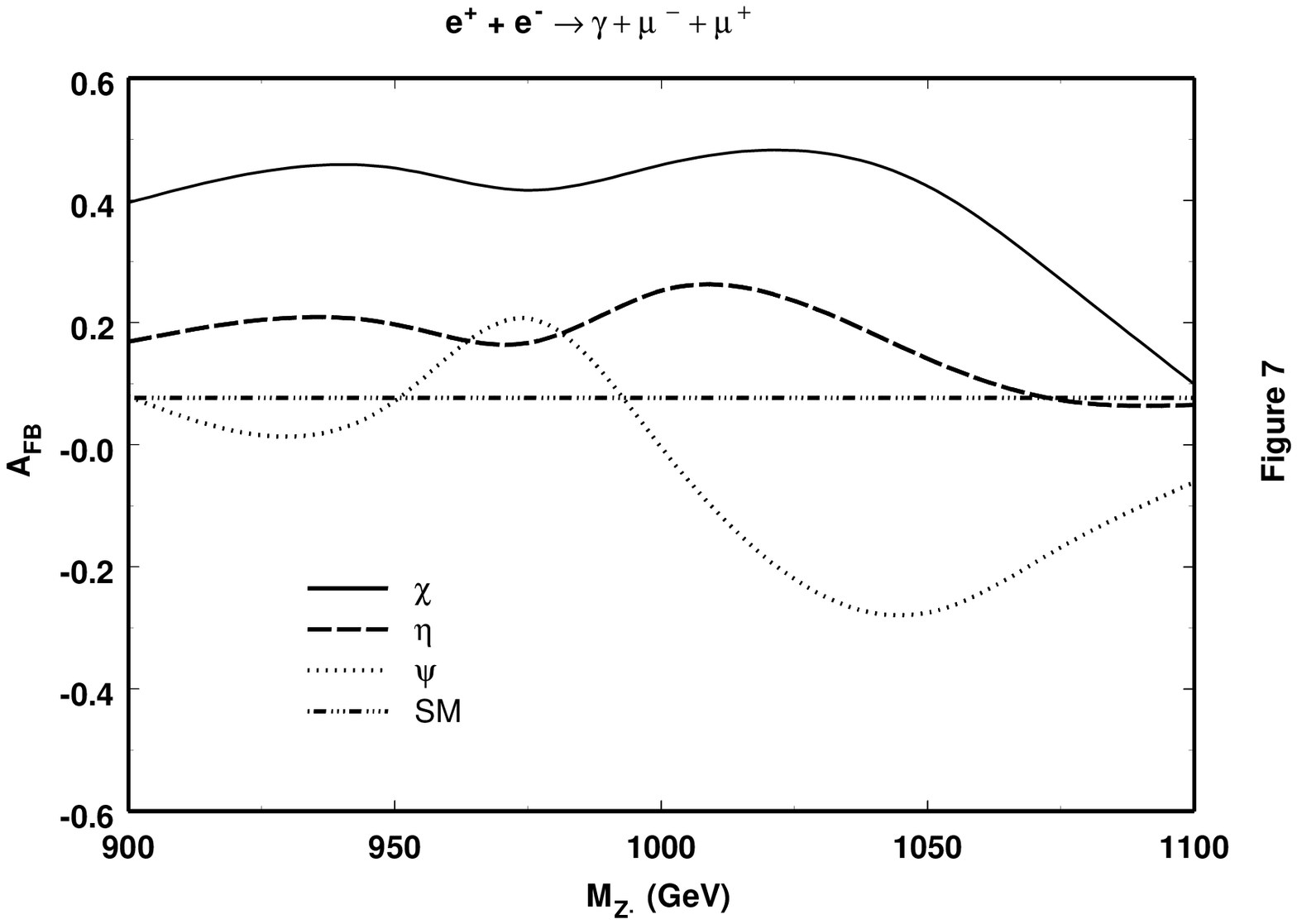}
       \end{center}
\caption{Forward-backward asymmetry as a function of $M_{Z^{\prime}}$ for the $\mu^-$ relative to the initial $e^-$ in $ e^+ e^- \longrightarrow  \gamma ~\mu^- ~\mu^+$ for the $ \chi , \eta$ and $\psi$ and standard models at $ {\sqrt s}= 1$ TeV.}
\end{figure}

\section{Conclusions}\setcounter{equation}{0}

\par
In this paper we have shown that the energy distribution of a hard photon in $ e^+ e^- \longrightarrow  \gamma ~ f  \bar f $  can give a very clear indication of a new neutral gauge boson with mass $M_{Z^{\prime}} < \sqrt s $ using a simple kinematical expression (2), instead of the reconstructing the $Z^{\prime}$ mass from the final fermions. We have simulated the finite resolution of detectors for final state photons and leptons and found that the hard photon energy distribution is practically not distorted. The reconstructed $M_{l^+ l^-}$ invariant mass distribution is flatter and its peak moved from the real $Z^{\prime}$ mass value. Analyzing hadronization effects for the channel $ e^+ e^- \longrightarrow Z^{\prime} \longrightarrow  \gamma ~ q \bar q $, it is obtained that the quark hadronization introduces experimental uncertainties, strongly modifying the distribution that are not present in the hard photon distribution. Analyzing the hard photon distribution it is possible also to sum over all
final fermions contributions, increasing substantially the  $Z^{\prime}$ mass statistic and its resolution.
\par
The final muons backward-forward asymmetry can be used to establish the relevant theoretical origin of a new possible $Z^{\prime}$.
\par
 The hard photon energy distribution approach could also be very useful in the case of more than one new neutral gauge boson with mass smaller than $\sqrt s$. Two or more peaks in the hard photon energy distribution will indicate two or more new neutral gauge bosons, without the need to tune the accelerator energy to the resonant values.  This analysis can be applied to any extended model with  extra neutral gauge bosons and to any future high energy lepton colliders such as the NLC, Tesla or CLIC. 

\par
\vskip 1cm
{\it Acknowledgments:} This work was partially supported by the

following Brazilian agencies: CNPq and FAPERJ.


\begin{thebibliography}{ABC}
\bibitem{ASS}J.~Hewett and T.~Rizzo, Phys. Rep. {\bf 183}, 193 (1989); A.~Leike,
Phys.\ Rep.\ {\bf 317}, 143 (1999).
\bibitem{JCP} J.~C.~Pati and A.~Salam, Phys. Rev. D {\bf 10}, 275 (1974); R.~N.~Mohapatra and J.~C.~Pati, Phys. Rev. D {\bf 11}, 566 (1975); G.~Senjanovi\u c and R.~N.~Mohapatra, Phys. Rev. D {\bf 12}, 1502 (1975); R.~N.~Mohapatra and R.~E.~Marshak, Phys. Lett. B {\bf91}, 222 (1980). An extensive list of references can be found in  R.~N.~Mohapatra and P.~B.~Pal, "Massive Neutrinos in Physics and Astrophysics", World Scientific, Singapore, 1998.
\bibitem{CVE} M.~Cveti\u c and S.~Godfrey, hep-ph/9504216; M.~Dittmar, A.~S.~Nicollerat and A.~Djouadi, Phys.\ Lett.\ B {\bf 583}, 111 (2004). 
\bibitem {GCV} G.~Cveti\u c and C.~S.~Kim, Phys. Lett. B {\bf 461}, 248 (1999).
\bibitem{DJO} A.~Djouadi, J.~Ng and T.~G.~Rizzo in: Electroweak Symmetry Breaking
and Beyond the Standard Model. Ed. T.~Barklow, S.~Dawson, H.~E.~Haber 
and S.~Siegrist, World Scientific, Singapore. hep-ph/ 9504210;
P.~Langacker, M.~Luo and A.~K.~Mann, Rev. Mod. Phys. {\bf64}, 87 (1992).
\bibitem{NLC} T.~Abe {\it et al.} [American Linear Collider Working Group Collaboration],
in {\it Proc. of the APS/DPF/DPB Summer Study on the Future of Particle Physics (Snowmass 2001) } ed. N.~Graf, hep-ex/0106055.
\bibitem{PDG} Particle Data Group, Phys. Lett. B {\bf 592}, 1 (2004).
\bibitem{APE} T. Appelquist, B.A. Dobrescu and A.R. Hopper, Phys. Rev. D { \bf 68}, 035012 (2003).
\bibitem{CAR} M. Carena, A. de Gouvea, A. Freitas and M. Schmitt, Phys. Rev. D {\bf 68}, 113007 (2003).
\bibitem{FRE} A.~Freitas, Phys. Rev. D {\bf 70}, 015008 (2004).
\bibitem{NOS} F.~M.~L.~Almeida, Y.~A.~Coutinho, J.~A.~Martins Sim\~oes and M.~A.~B.~do Vale,
Phys.\ Lett.\ B {\bf 494}, 273 (2000); F.~M.~L.~Almeida, Y.~A.~Coutinho, J.~A.~Martins Sim\~oes, S.~Wulck and M.~A.~B.~do Vale, Eur.\ Phys.\ J.\ C {\bf 30}, 327 (2003).
\bibitem{NCL}  G.~A. Blair,[Physics at Tesla], hep-ex/0104044, and references therein; S.~Kuhlman {\it et al.} [NLC ZDR Design Group and NLC Physics Working Group Collaboration], hep-ex/9605011.
\bibitem {HEP} A.~Pukhov, E.~Boos, M.~Dubinin, V.~Edneral, V.~Ilyin, D.~
Kovalenko, A.~Kryukov, V.~Savrin, S.~Shichanin and A.~Semenov, 
"CompHEP"- a package for evaluation of Feynman diagrams and integration over 
multi-particle phase space. Preprint INP MSU 98-41/542, hep-ph/9908288.
\bibitem {SME} R.~Settles, H.~Spiesberger and W.~Wiedenmann, Smear version 3.02; http://www.desy.de/~hspiesb/smear.html

\end{thebibliography}
\end{document}